\documentclass[prl,amsmath,amssymb,twocolumn, aps,superscriptaddress,floatfix,preprintnumbers,10pt, notitlepage]{revtex4-1}
\RequirePackage[switch,columnwise]{lineno}
\usepackage{graphicx}
\usepackage{amsmath, braket, amsfonts}
\usepackage{amssymb}
\usepackage{natbib}
\usepackage{sidecap}
\usepackage{bm, color, ulem}
\usepackage{comment}
\usepackage{xcolor}
\usepackage{tikz}
\usepackage{pgfplots}
\usepackage{setspace}
\usepackage{titlesec}
\pgfplotsset{compat=1.12}
\titleformat*{\section}{\normalsize\bfseries}

\usepackage{hyperref}
\hypersetup{
 pdfnewwindow=true, colorlinks=true,
 linkcolor=blue, anchorcolor=blue,
 citecolor=blue, filecolor=blue,
 menucolor=blue, urlcolor=blue} 

\begin{document}
\title{Hofstadter subband ferromagnetism and symmetry broken Chern insulators in twisted bilayer graphene}
\author{Yu Saito}
\affiliation{Department of Physics, University of California at Santa Barbara, Santa Barbara CA 93106, USA}
\affiliation{California NanoSystems Institute, University of California at Santa Barbara, Santa Barbara CA 93106, USA}
\author{Jingyuan Ge}
\affiliation{Department of Physics, University of California at Santa Barbara, Santa Barbara CA 93106, USA}
\author{Louk Rademaker}
\affiliation{Department of Theoretical Physics, University of Geneva, 1211 Geneva, Switzerland}
\author{Kenji Watanabe}
\affiliation{National Institute for Materials Science, 1-1 Namiki, Tsukuba 305-0044, Japan}
\author{Takashi Taniguchi}
\affiliation{National Institute for Materials Science, 1-1 Namiki, Tsukuba 305-0044, Japan}
\author{Dmitry A. Abanin}
\affiliation{Department of Theoretical Physics, University of Geneva, 1211 Geneva, Switzerland}
\author{Andrea F. Young}
\email[Electronic address:]{andrea@physics.ucsb.edu}
\affiliation{Department of Physics, University of California at Santa Barbara, Santa Barbara CA 93106, USA}
\date{\today}

\begin{abstract}
 \end{abstract}
\maketitle

\textbf{
In bilayer graphene rotationally faulted to $\bm{\theta\approx 1.1^\circ}$, interlayer tunneling and rotational misalignment conspire to create a pair of low energy flat bands\cite{bistritzer_moire_2011-1} that have been found to host a variety of insulating, superconducting, and magnetic states at partial filling\cite{cao_correlated_2018,cao_unconventional_2018,yankowitz_tuning_2019,lu_superconductors_2019,sharpe_emergent_2019,serlin_intrinsic_2020}. 
Most work to date has focused on the zero magnetic field phase diagram, with magnetic field used as a probe of the $\bm{B = 0}$ band structure.
Here, we show that twisted bilayer graphene (tBLG) in magnetic fields of several Tesla hosts a cascade of ferromagnetic Chern insulators with Chern number $\bm{|C| = 1, 2}$ and $\bm{3}$. 
We argue that the emergence of the Chern insulators is driven by the interplay of the moir\'e superlattice with the magnetic field, which endow the flat bands with a substructure of topologically nontrivial subbands characteristic of the Hofstadter butterfly\cite{hofstadter_energy_1976,thouless_quantized_1982}. 
The new phases can be accounted for in a Stoner picture\cite{nomura_quantum_2006} in which exchange interactions favor polarization into one or more spin- and valley-isospin flavors; in contrast to conventional quantum Hall ferromagnets, however, electrons polarize into between one and four copies of a single Hofstadter subband with Chern number $\bm{C = -1}$\cite{bistritzer_moire_2011-1,moon_energy_2012,zhang_landau_2019}. 
In the case of the $\bm{C = \pm 3}$ insulators in particular, magnetic field catalyzes a first order-like phase transition from the spin- and valley-unpolarized $\bm{B=0}$ state into the ferromagnetic state. 
Distinct from other moir\'e heterostructures\cite{dean_hofstadters_2013,ponomarenko_cloning_2013,hunt_massive_2013}, tBLG realizes the strong-lattice limit of the Hofstadter problem and hosts Coulomb interactions that are comparable to the full bandwidth $\bm{W}$ and are consequently much stronger than the width of the individual Hofstadter subbands. In our experimental data, the dominance of Coulomb interactions manifests through the  appearance of Chern insulating states with spontaneously broken superlattice symmetry at half filling of a $\bm{C=-2}$ subband\cite{wang_evidence_2015,spanton_observation_2018}. Our experiments show that tBLG may be an ideal venue to explore the strong interaction limit within partially filled Hofstadter bands.} 

\begin{figure*}[ht!]
\includegraphics[width= 178mm]{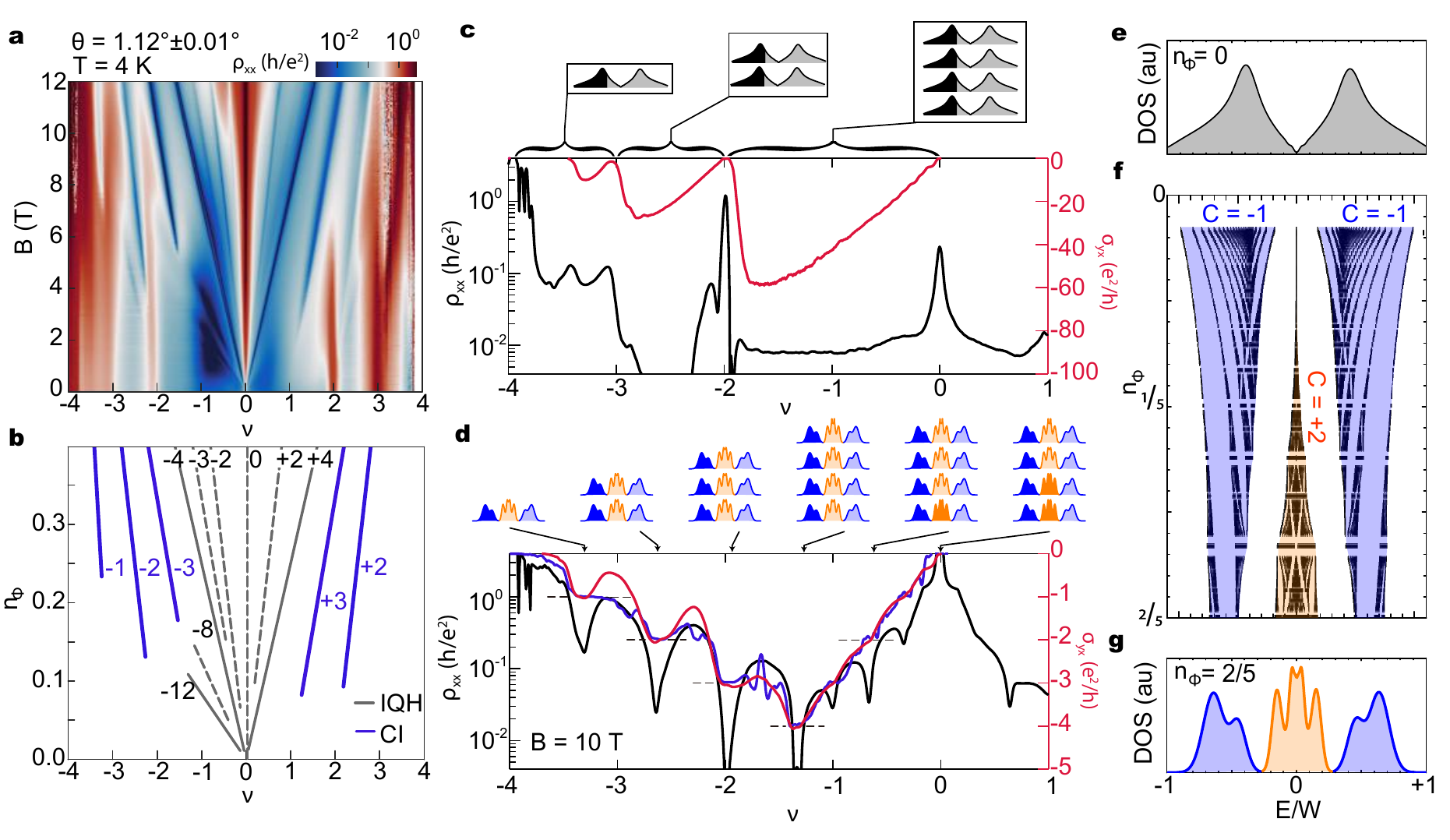}
\caption{\textbf{Hofstadter subband ferromagnetism.}
\textbf{a}, Longitudinal resistivity $\rho_\mathrm{xx}$ as a function of superlattice filling factor $\nu$ and magnetic field $B$ at 4 K.
\textbf{b}, Schematic of the Chern insulator structure observed in panel a, showing gapped states that following linear trajectories $\nu = tn_\mathrm{\Phi} + s$ where $n_\Phi$ denotes the magnetic flux quanta per superlattice unit cell area. We estimate $n_\phi=1$ for $B=30.1\pm0.1$ T.  
Two trajectory classes are distinguished by color corresponding to $s = 0, t \in \mathbb{Z}$ integer quantum Hall states (gray) and  $s, t \in \mathbb{Z}, s \neq 0$ Hofstadter Chern insulators (blue). 
\textbf{c}, $\rho_\mathrm{xx}$ (black) and $\sigma_\mathrm{yx}$ (red) as a function of $\nu$ at 0.6 K at 0 T ($\rho_\mathrm{xx}$) and 0.6 T ($\sigma_\mathrm{yx}$). Schematic images show the pattern of flavor filling consistent with observed low magnetic field quantum oscillations (see supplementary information).
\textbf{d}, $\rho_\mathrm{xx}$ at $T = 4$ K (black) and $\sigma_\mathrm{yx}$ at 4 K (red) and 10 mK (blue) as a function of $\nu$ at $B$ = 10 T. Dashed line shows integer denote integer multiples of $e^2/h$. 
Schematic images show sequence of filling of Hofstadter subbands. \textbf{e}, Calculated density of states (DOS) of the valence and conduction flat bands at zero magnetic field.
\textbf{f}, Calculated Hofstadter spectra of tBLG. Blue and Yellow region correspond to Chern bands with total Chern number of $C = -1$ and $+2$, respectively. (see supplementary information for details.)
\textbf{g}, Calculated DOS at $n_{\Phi} = 2/5$.}
\label{fig:1}
\end{figure*}

The energy spectrum of a two dimensional electron subjected simultaneously to a magnetic field and periodic potential is described by a fractal structure known as the Hofstadter butterfly\cite{hofstadter_energy_1976}. 
For a system with $n$ electronic bands at zero magnetic field, the Hofstadter spectrum at a magnetic flux per unit cell of $\Phi/\Phi_0=p/q$ hosts $n\times q$ `subbands' (here $\Phi_0=h/e$ is the magnetic flux quantum and $p$ and $q$ are integers with  greatest common divisor 1).  Remarkably, these subbands are each characterized by a nonzero, integer-valued topological index, known as a Chern number $C$, which describes their contribution to the quantized Hall conductivity when an integer number of subbands are filled\cite{thouless_quantized_1982}.  
Owing to their high sample quality and the large area of the superlattice, moir\'e van der Waals heterostructures have provided a versatile materials system for exploring the physics of the Hofstadter subbands\cite{dean_hofstadters_2013,ponomarenko_cloning_2013,hunt_massive_2013}, including the observation of correlation driven Chern insulators at fractional filling of a Hofstadter subband that either spontaneously break the superlattice symmetry\cite{wang_evidence_2015} or fractionalize electrons into anyons\cite{spanton_observation_2018}.  In typical moir\'e heterostructures of graphene and hexagonal boron nitride (hBN), however, accessing the Hofstadter regime experimentally requires strong magnetic fields.  This is tied to the fact that the superlattice potential induced by the moir\'e pattern is weak, and provides only a small perturbation to the conventional Landau levels.  In this limit, Hofstadter subbands form within a single Landau level, and are restricted to an energy bandwidth per Landau level proportional to $W\propto \exp\left(-\frac{\pi}{2}\frac{\Phi_0}{\Phi}\right)$\cite{dean_fractional_2020}. Achieving separations between subbands that exceed the energy scales related to disorder or temperature thus requires a significant fraction of the magnetic flux quantum to be threaded through each superlattice unit cell.

Electrons in the tBLG flat bands are strongly localized and should consequently realize a strong-lattice limit of the Hofstadter butterfly in a finite magnetic field. In this limit, topologically nontrivial subbands may be separated by a significant fraction of the total bandwidth even for $\Phi \ll \Phi_0$\cite{bistritzer_moire_2011,moon_energy_2012,zhang_landau_2019}.  
However, discussions in the literature of the magnetotransport properties of tBLG  have focused on relating the observed magnetoresistance features to the enigmatic $B=0$ phase diagram, without regard for the interplay of the magnetic flux with the periodic potential. 
Here, we show that magnetic fields can indeed drive new symmetry breaking transitions, producing magnetoresistance features at low magnetic fields that are not correlated with the realized $B=0$ ground states.  
Fig. \ref{fig:1}a shows longitudinal resistivity ($\rho_\mathrm{xx}$) data from a nominally h-BN-unaligned $\theta=1.12^\circ$ tBLG device (see Fig. \ref{fig:device_contact}) as a function of magnetic field ($B$). We present data at 4 K to suppress all but the most energetically robust features; lower temperature $\rho_\mathrm{xx}$ data and Hall conductivity ($\sigma_\mathrm{yx}$) data are available in Figs. \ref{fig:Full LL supple} and \ref{fig:Hall_landaufan}. 
At $B = 0$, our device shows familiar features of flat band tBLG: correlated insulators or resistivity peaks at $\nu=\pm2$, $\pm3$, and $0$ and a robust superconducting state for $\nu\lesssim-2$ (see Figs. \ref{fig:1}c and \ref{fig:SC}). Here $\nu$ indicates the number of electrons per superlattice unit cell.  
However, at $B\gtrsim 5$ T (for $\nu<0$) and $B\gtrsim2$ T (for $\nu>0$), this behavior abruptly changes, 
giving way to a series of Chern insulators, indicated in blue in Fig. \ref{fig:1}b. They are characterized by $\rho_\mathrm{xx}$ minima and integer quantized Hall conductivity that increases in magnitude as the band is filled.

\begin{figure}[ht!]
\centering
\includegraphics[width= 89mm]{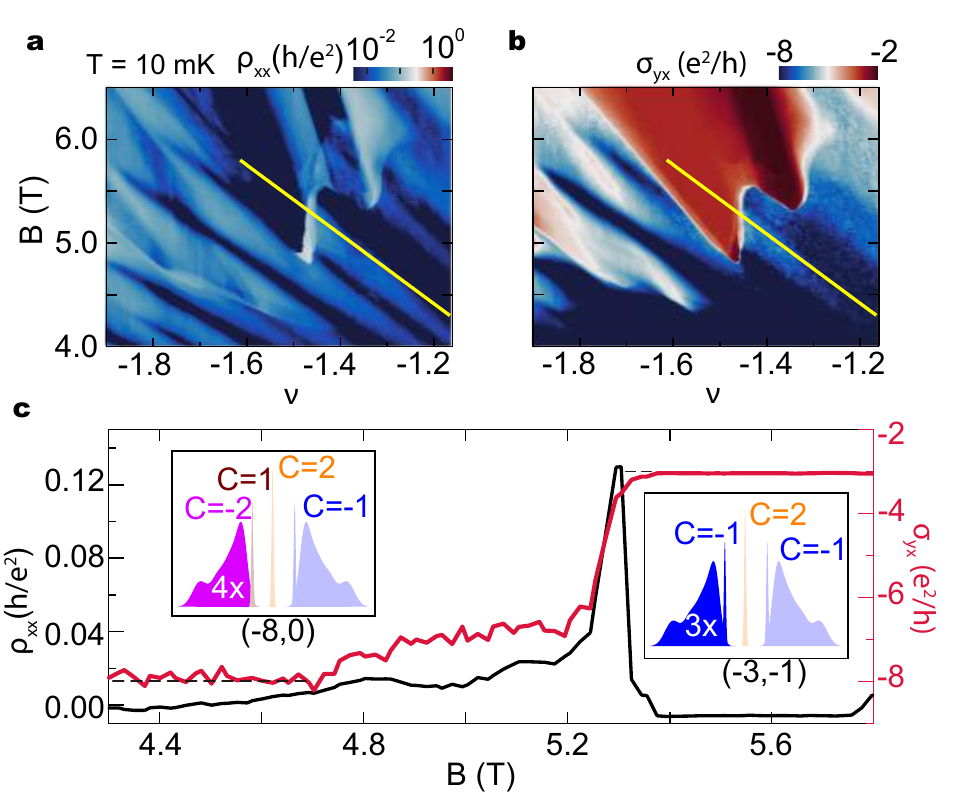}
 \caption{\textbf{Magnetic field-driven symmetry breaking transition to a $C = -3$ Chern insulator.}
\textbf{a}, $\rho_\mathrm{xx}$  and \textbf{b}, $\sigma_\mathrm{yx}$ as a function of $\nu$ and $B$ between 4 and 6.5 T, measured at a nominal temperature of 10 mK.
\textbf{c}, Line cuts of $\rho_\mathrm{xx}$ (black) and $\sigma_\mathrm{yx}$ (red) along yellow lines in \textbf{a} and \textbf{b}. 
Dashed lines correspond to $-3e^2/h$ and $-8e^2/h$. Cartoons depict inferred band fillings for $(t, s) = (-8, 0)$ and $(-3, 1)$. At $(-3,-1)$, three $C=-1$ subbands are filled and one is empty. At $(-8,0)$ all four flavors are equally occupied, filling four copies of a $C = -2$ band including the entire $C=-1$ band except for the highest energy state associated with a single, strain split Landau level\cite{zhang_landau_2019}.}
\label{fig:2}
\end{figure}

Within the Hofstadter picture, gapped states follow linear trajectories according to a Diophantine equation, $\nu = tn_\mathrm{\Phi} + s$ ($s, t \in \mathbb{Z}$) in the space of $\nu$ and  $n_\mathrm{\Phi} \equiv \Phi/\Phi_0$\cite{wannier_result_1978}, the magnetic flux per moir\'e unit cell. Here $s$ is the Bloch band filling index, which encodes the number of electrons bound to each lattice unit cell, while $t$ is the total Chern number associated with a given gap, and consequently the quantized Hall conductivity\cite{streda_quantised_1982,thouless_quantized_1982}.  
Starting from $\nu=-4$ and moving into the flat band, at $n_\mathrm{\Phi} = 0.3$ the observed sequence follows $(t,s)=(-1,-3), (-2,-2), (-3,-1)$ and $(-4,0)$. 
This pattern is consistent with sequentially filling four bands with Chern number $C=-1$, each of which binds a single electron per moir\'e unit cell.

The observed transition between low and high magnetic field regimes can be understood from the evolution of the tBLG flat bands in a magnetic field.  
At $B=0$, tBLG hosts two low-energy bands (for a given spin and valley flavor) connected by two Dirac points that remain gapless absent symmetry breaking by substrate potentials\cite{po_origin_2018,kang_symmetry_2018,koshino_maximally_2018}.  
A key feature of the $B=0$ band structure is the triangular form of the density of states, shown in Fig.~\ref{fig:1}e.  
It was recently argued\cite{zondiner_cascade_2020} that exchange interactions within such a band lead to a peculiar form of Stoner ferromagnetism in which exchange interactions favor collective states that are symmetric between two, three, or four isospin flavors. In our device, the low $B$ quantum oscillations (see Fig. \ref{fig:low field fan}) are qualitatively consistent with this mechanism, showing signatures of Dirac fermions with fully broken flavor symmetry for $\nu<-3$, twofold symmetry for $-3<\nu<-2$, and fourfold symmetry between $-2<\nu<0$ (see insets to Fig. \ref{fig:1}c and \ref{fig:low field fan}) 
Notably, in our device, and indeed in the majority of published transport data in tBLG devices, symmetry breaking is absent at $\nu=\pm1$, as evidenced by the absence of low-$B$ quantum oscillations near those fillings. This effect is not explained within the model of Reference \onlinecite{zondiner_cascade_2020}, but may arise due to the reconstruction of the low energy bands that is expected to occur at partial fillings due to the Coulomb interactions\cite{guinea_electrostatic_2018}.

An out of plane magnetic field significantly reconstructs the low energy bands.  Figure \ref{fig:1}f shows the evolution of electronic structure as a function of magnetic field, while Fig. \ref{fig:1}g shows the calculated density of states (DOS) at $n_\mathrm{\Phi}=2/5$, corresponding to $B=12$ T for our device. At low magnetic fields, the magnetic field reconstructs the two flat bands into three groups of subbands (for each spin and valley) having net Chern numbers $-1$, $+2$, and $-1$ respectively. This structure is generic to all tBLGs, with the $C=2$ subband originating in the zero energy Landau level in the two constituent graphene monolayers and the $C=-1$ bands ensuring that the combined bands maintain net Chern number of zero.  Even in moderate fields the DOS is characterized by three effective subbands that contrast to the two bands at $B=0$. The cascade of Chern insulators is well explained by the sequential filling of four spin and valley-projected copies of the lower $C = -1$ subbands, as illustrated in Fig \ref{fig:1}d and emerges from a simplified analysis of a Stoner model that accounts for the magnetic-field induced change in bandwidth (see Supplementary information and Fig. \ref{Fig:SupplTheory2}).
This picture is mirrored for electron doping, and is also consistent with the Landau fan originating from the CNP, which shows the strongest features at $(t,s)=(\pm4,0),(\pm2,0)$, and $(0,0)$, consistent with successive fillings of the central $C=2$ subband.

\begin{figure*}[ht!]
\includegraphics[width= 130mm]{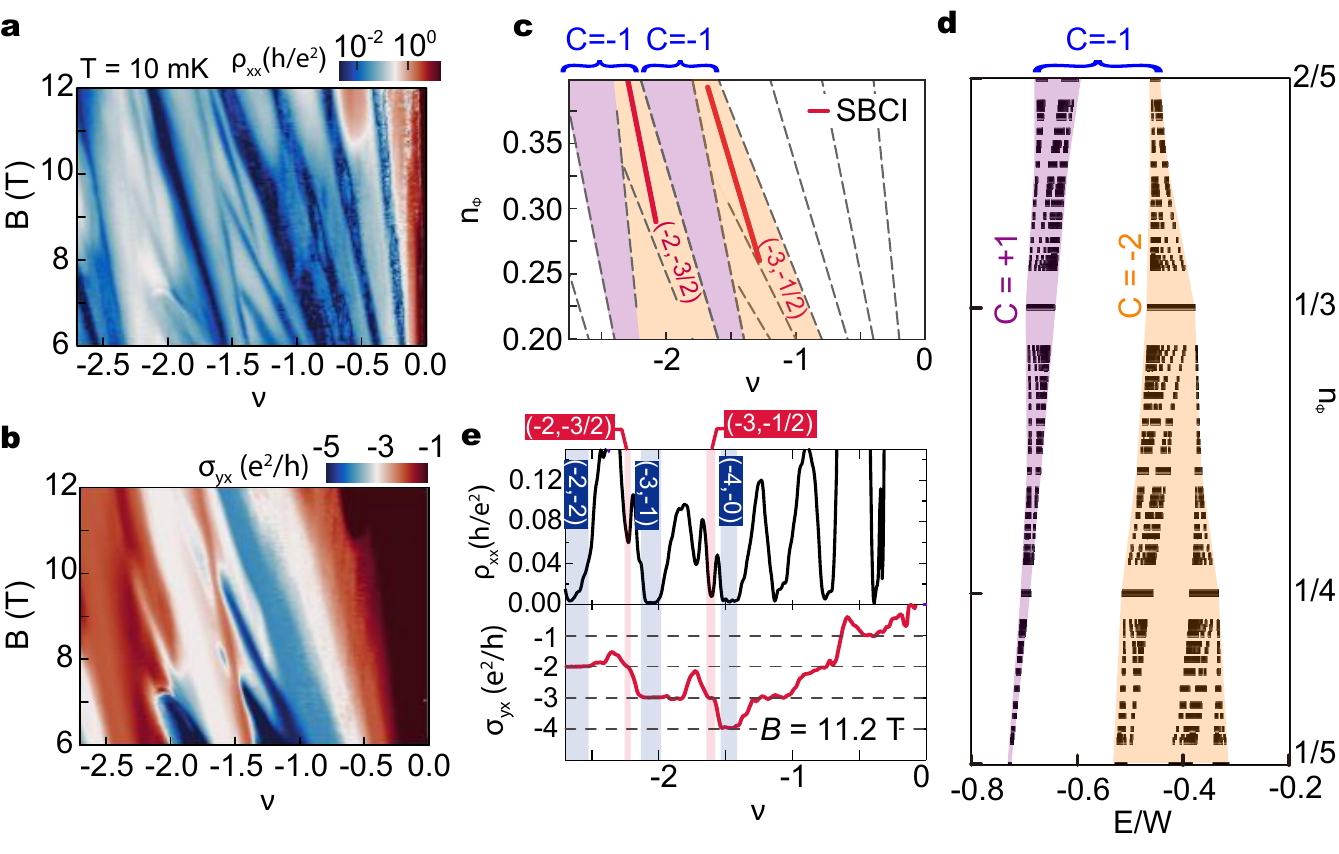}
 \caption{\textbf{Symmetry broken Chern insulators.}
 \textbf{a}, 
 $\rho_\mathrm{xx}$ and \textbf{b}, $\sigma_\mathrm{yx}$ as a function of $\nu$ and $B$ in a range of 6-12 T at a nominal temperature of 10 mK.
 \textbf{c}, Schematic of observed Chern insulator states panels \textbf{a, b}. 
 Red lines correspond to symmetry broken Chern insulator (SBCI) with $(t, s) = (-2, -3/2)$ and $(-3, -1/2)$.
 \textbf{d},  Calculated Hofstadter energy spectrum in the regime of \textbf{c}, focusing on the low-energy $C=-1$ band. The two primary subbands color coded, and the equivalent extent in density is shown in panel \textbf{c} for two copies the band.
 \textbf{e}, Line cuts of $\rho_\mathrm{xx}$ and $\sigma_\mathrm{yx}$ at 11.2 T with several Chern insulator states labeled. 
}
\label{fig:3}
\end{figure*}

The most striking features of the Chern insulator cascade are the states at $(t,s)=(\pm 3, \pm 1)$(Fig. \ref{fig:2} and \ref{fig:p_landaufan}), which are not correlated with a set of low-magnetic field  quantum oscillations (Fig. \ref{fig:low field fan}).  
Indeed, low temperature measurements show that these states emerge without warning in the midst of well formed four-component quantum oscillations originating from the charge neutrality point, consistent with previous reports of orbital magnetic states at finite $B$ \cite{lu_superconductors_2019,stepanov_interplay_2019}.  The sudden phase transition (Figs. \ref{fig:2}a and b) and sudden appearance of an activation gap (Fig. \ref{fig:gap_temp}) can be understood within the context of Stoner ferromagnetism that accounts for the Chern character of the $C=-1$ Hofstadter subband.  Within this picture, symmetry breaking occurs when the density of states $\mathrm{DOS}>1/U$, where $U$ is the exchange interaction strength. 
As the magnetic field increases, both the bandwidth of and total number of states within the $C = -1$ subband decrease; numerically, however, this leads to an increase in average DOS (see Fig. \ref{Fig:SupplTheoryDOS} and supplementary information), favoring symmetry breaking above some critical $B$. Here, we did not observe any hysteric behavior in this transition.
The nature of this first order-like transition is illustrated in Fig. 2c, which shows a line cut connecting the $(-3,-1)$ and $(-8,0)$ states. At $(-3,-1)$, three $C=-1$ subbands are completely filled and one empty; while at $(-8,0)$ all four flavors are equally occupied, filling four copies of a $C = -2$ band that includes the entire $C=-1$ band except for the highest energy state associated with a single, strain split Landau level in each flavor\cite{zhang_landau_2019}.

In one view, the Chern insulator states can be seen as competing ground states in twisted bilayer graphene at integer band filling and $B=0$\cite{xie_nature_2020}. It is known that small perturbations such as an aligned hBN substrate can favor Chern insulators at $B=0$, for instance at $\nu=3$\cite{serlin_intrinsic_2020}. Even absent this splitting, however, a retrospective review shows that nearly all published data containing magnetotransport data show a feature likely associated with the transition we report here to one or more of the Chern insulator states we observe here\cite{cao_unconventional_2018,yankowitz_tuning_2019,lu_superconductors_2019,stepanov_interplay_2019,uri_mapping_2019}.  These transitions can be generically understood as reflecting the finite orbital moment of the Chern insulator states, which lowers their energy relative to competing $C=0$ insulators or semimetallic ground states in a finite magnetic field. 
However, we prefer to analyze the observed ferromagnetic Chern insulator states as generalizations of quantum Hall ferromagnets in which the Landau levels are replaced by $C=-1$ subbands of the tBLG Hofstadter spectrum. 
We note that just as in graphene Landau levels\cite{dean_fractional_2020}, each Chern ferromagnet  corresponds to polarization within the combined space of spins and valleys with the precise isospin ordering set by the interplay of higher order effects including the Zeeman effect and the anisotropy of the Coulomb interactions themselves.  
Interestingly, these anisotropies may differ within the Chern bands as compared to both Landau levels and the $B=0$ flat bands, possibly favoring different broken symmetries at integer fillings. 

This approach moreover has the advantage of providing a quantitative description of the electronic substructure of the $C=-1$ band at high magnetic fields. Figures \ref{fig:3}a and b display measured $\rho_\mathrm{xx}$ and $\sigma_\mathrm{yx}$, respectively, from 6 to 12 T in a density region corresponding to occupying the third and fourth copies of the lower $C=-1$ bands.  
The low disorder in our graphite gated sample\cite{zibrov_tunable_2017} allows us to resolve a number of smaller energy gaps characterized within the $C=-1$ band. 
First, we resolve the next level of subband structure within the $C=-1$ subband, which is composed of well separated clusters of subbands with net $C=+1$ and $C=-2$ (see Figs. \ref{fig:3}c-d). These features are repeated in experimental data, consistent with the scenario of sequential filling of the overall $C=-1$ band described above. 
In addition, we also observe robust states showing  $(t, s) = (-3, -1/2)$ and $(-2, -3/2)$ at half filling of the $C=-2$ band (highlighted in red in Fig. \ref{fig:3}c). These states show quantized $\sigma_\mathrm{yx}=t e^2/h$ as expected from the measured slope $t$, and an activation gap of around 1 K (Fig. \ref{fig:3}e, \ref{fig:sbci} and \ref{fig:sbci_gap}). Similar states are observed at positive fillings as well (Fig.~\ref{fig:sbci}).  

The quantum number $s$ encodes the total number of electrons bound to a given lattice site at fixed magnetic flux, and is an integer for all states that can be described within a noninteracting Hofstadter butterfly. The simplest mechanism that allows this number to be fractional is for the superlattice symmetry to break, and we associated the observation of states with half integer $s$ to the formation of symmetry broken Chern insulators, in which electronic interactions within a $C=-2$ Hofstadter subband drive spontaneous doubling of the unit cell. 
While such states were previously observed in hBN-aligned graphene\cite{wang_evidence_2015,spanton_observation_2018}, they typically appear only at magnetic fields in excess of 16 T, more than twice as high as in tBLG.  
The relative strength of the SBCI states can be related to the strong lattice limit of the Hofstadter band structure realized in tBLG. Specifically, within the Hofstadter subbands of a weak-superlattice system the Coulomb interaction scale is never stronger than $e^2/\ell_B$ (here $\ell_B=\sqrt{\hbar/eB}$ is the magnetic length), which vanishes at low $B$. 
In contrast, in tBLG and other moir\'e flat band systems the magnetic-field independent moir\'e wavelength sets the scale for interactions. Interactions can drive the formation of correlated states at partial filling of Hofsatdter subbands in the low field limit, just as they drive correlated states at $B=0$. 
Our experiment suggests that other exotic ground states at fractional fillings of Chern bands may be accessible in moir\'e flat band systems, potentially even at $B=0$ in the presence of appropriate time-reversal symmetry breaking order.
\textit{Note:} During the preparation of this work we became aware of two additional reports reporting some of the same observations\cite{nuckolls_strongly_2020,wu_chern_2020}.


\let\oldaddcontentsline\addcontentsline
\renewcommand{\addcontentsline}[3]{}
\section*{Methods}
\noindent\textbf{Device fabrication}\\ The tBLG device used in this study were fabricated using a ``cut-and-stack'' technique described in Ref. \onlinecite{saito_independent_2020}, and indeed the device is the same as device \#5 in that paper. Prior to stacking, we first cut graphene into two pieces using AFM to prevent the unintentional strain in tearing graphene. We used a poly(bisphenol A carbonate) (PC)/polydimethylsiloxane (PDMS) stamp mounted on a glass slide to pick up a 40-nm-thick hBN flake at 90$-$110$^\circ$C, and carefully pick up the 1st half of a pre-cut graphene piece, rotate the sample stage and pick up the 2nd half of the graphene at 25 $^\circ$C using the hBN flake. We rotated the graphene pieces manually by a twist angle of about 1.2$^\circ-$1.3$^\circ$ before the second pickup. Finally, the 3-layer stack (hBN-tBLG) is transferred onto another stack (40-nm-thick hBN-graphite) containing the bottom gate and gate dielectric, which is prepared in advance by the same dry transfer process and cleaned by the typical solvent wash using chloroform, acetone, methanol and IPA followed by vacuum annealing (400$^\circ$C for 8 hours) to remove the residue of PC film on the hBN surface. 
Electrical connections to the tBLG were made by CHF$_3$/O$_3$ etching and deposition of the Cr/Pd/Au (2/15/180 nm) metal edge-contacts\cite{wang_one-dimensional_2013}. 

\noindent\textbf{Transport measurements}\\
The insulating nature of the $\nu=2,3$ and $4$ states appears to be correlated with poor contact resistances $R>1$ M$\Omega$ for $\nu\gtrsim 3$. We thus focus the bulk of our analysis on negative filling. All transport measurements in this study were carried out in a top-loading dilution refrigerator (Bluefors LD400) with a nominal base temperature of 10 mK, which is equipped with a 14 T superconducting magnet and heavy RF and audio frequency filtering with a cutoff frequency of $\sim$ 10 kHz. The temperature dependent measurements were done by controlling the temperature using a heater mounted on a mixing chamber plate. Standard low frequency lock-in techniques with Stanford Research SR860 amplifiers were used to measure the resistance with an excitation current of 2$-$3 nA at a frequency of 17.777 Hz. 

The twist angle $\theta$ is determined from the values of charge carrier density at which the insulating states at $n_{\nu = \pm 2}$ are observed, following $n_{\nu = \pm 2} = \pm 4 \theta^2/\sqrt{3}a^2$ , where $a$= 0.246 nm is the lattice constant of graphene.
 
\vspace{5pt}

\let\addcontentsline\oldaddcontentsline
\let\oldaddcontentsline\addcontentsline
\renewcommand{\addcontentsline}[3]{}

\section*{Acknowledgments}
\vspace{-12pt}
\noindent
The authors acknowledge discussions with E. Berg, I. Protopopov, T. Senthil, and M. Zaletel.
The experimental work was primarily supported by the ARO under W911NF-17-1-0323.
Y.S. acknowledges the support of the Elings Prize Fellowship from the California NanoSystems Institute at University of California, Santa Barbara.
K.W. and T.T. acknowledge support from the Elemental Strategy Initiative conducted by the MEXT, Japan and the CREST (JPMJCR15F3), JST.
AFY acknowledges the support of the David and Lucille Packard Foundation under award 2016-65145.
L.R. was supported by the Swiss National Science Foundation via an Ambizione grant. D.A. acknowledges the support of the Swiss National Science Foundation. 
\let\addcontentsline\oldaddcontentsline

\let\oldaddcontentsline\addcontentsline
\renewcommand{\addcontentsline}[3]{}

\section*{Author Contributions}
\vspace{-12pt}
\noindent
Y.S. and J.G. fabricated tBLG devices. Y.S. performed the measurements and analyzed the data. L.R. and D.A. performed the theoretical calculations. Y.S. and A.F.Y wrote the paper with input from L.R. and D.A. T.T. and K.W. grew the hBN crystals.
\let\addcontentsline\oldaddcontentsline

\let\oldaddcontentsline\addcontentsline
\renewcommand{\addcontentsline}[3]{}
\section*{Competing interests}
\vspace{-12pt}
\noindent
The authors declare no competing financial interests.
\let\addcontentsline\oldaddcontentsline

\let\oldaddcontentsline\addcontentsline
\renewcommand{\addcontentsline}[3]{}
\bibliographystyle{unsrt}
\bibliography{references}
\let\addcontentsline\oldaddcontentsline

\clearpage

\pagebreak
\widetext
\begin{center}
\textbf{\large Supplementary Information for \\Hofstadter subband ferromagnetism and symmetry broken Chern insulators in twisted bilayer graphene}\\
\vspace{10pt}
Yu Saito, Jingyuan Ge, Louk Rademaker, Kenji Watanabe, Takashi Taniguchi, Dmitry A. Abanin and Andrea F. Young\\
\vspace{10pt}
Correspondence to: andrea@physics.ucsb.edu\\
\vspace{10pt}
\end{center}
\renewcommand{\thefigure}{S\arabic{figure}}
\renewcommand{\thesubsection}{S\arabic{subsection}}
\setcounter{secnumdepth}{2}
\renewcommand{\theequation}{S\arabic{equation}}
\renewcommand{\thetable}{S\arabic{table}}
\setcounter{figure}{0}
\setcounter{equation}{0}
\onecolumngrid


\begin{figure*}[ht!]
\includegraphics[width=\columnwidth]{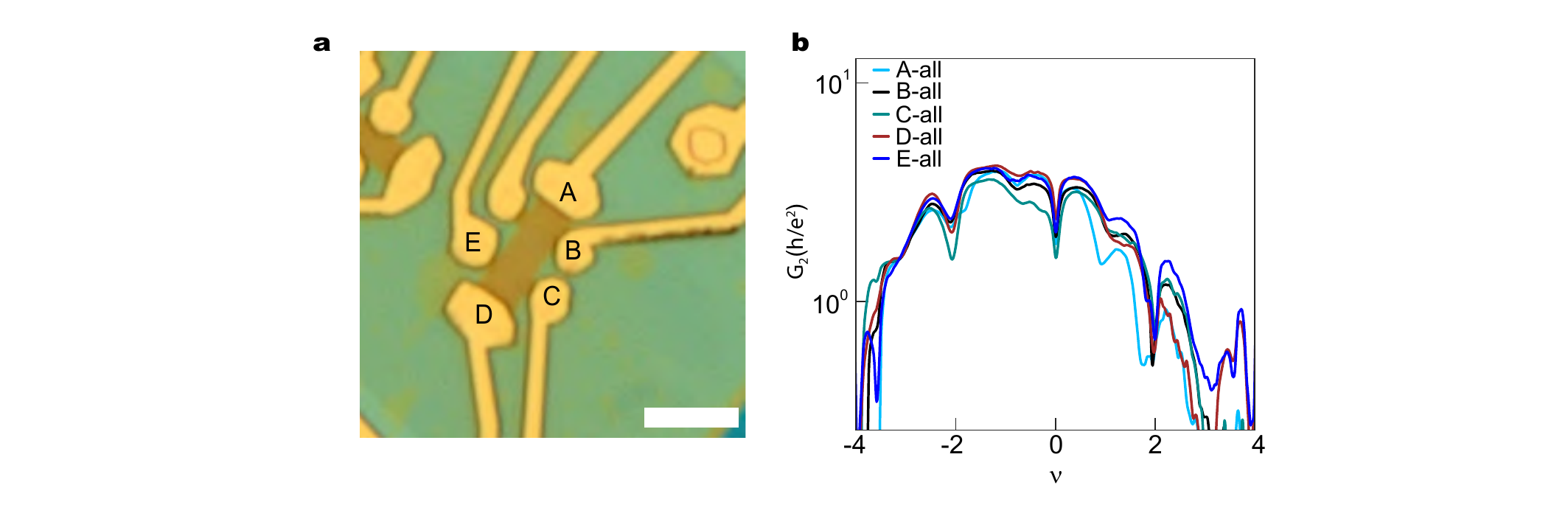}
 \caption{\textbf{Device characterization} 
 \textbf{a}, Optical microscope images of a device used in this study. Scale bar 5 $\mu$m. 
 \textbf{b}, Two terminal conductance $G_2$ across multiple contacts at 4 K showing high degree of uniformity. ``All" denotes all other contacts.}
\label{fig:device_contact}
\end{figure*}

\begin{figure*}[ht!]
\includegraphics[width= 7.2in]{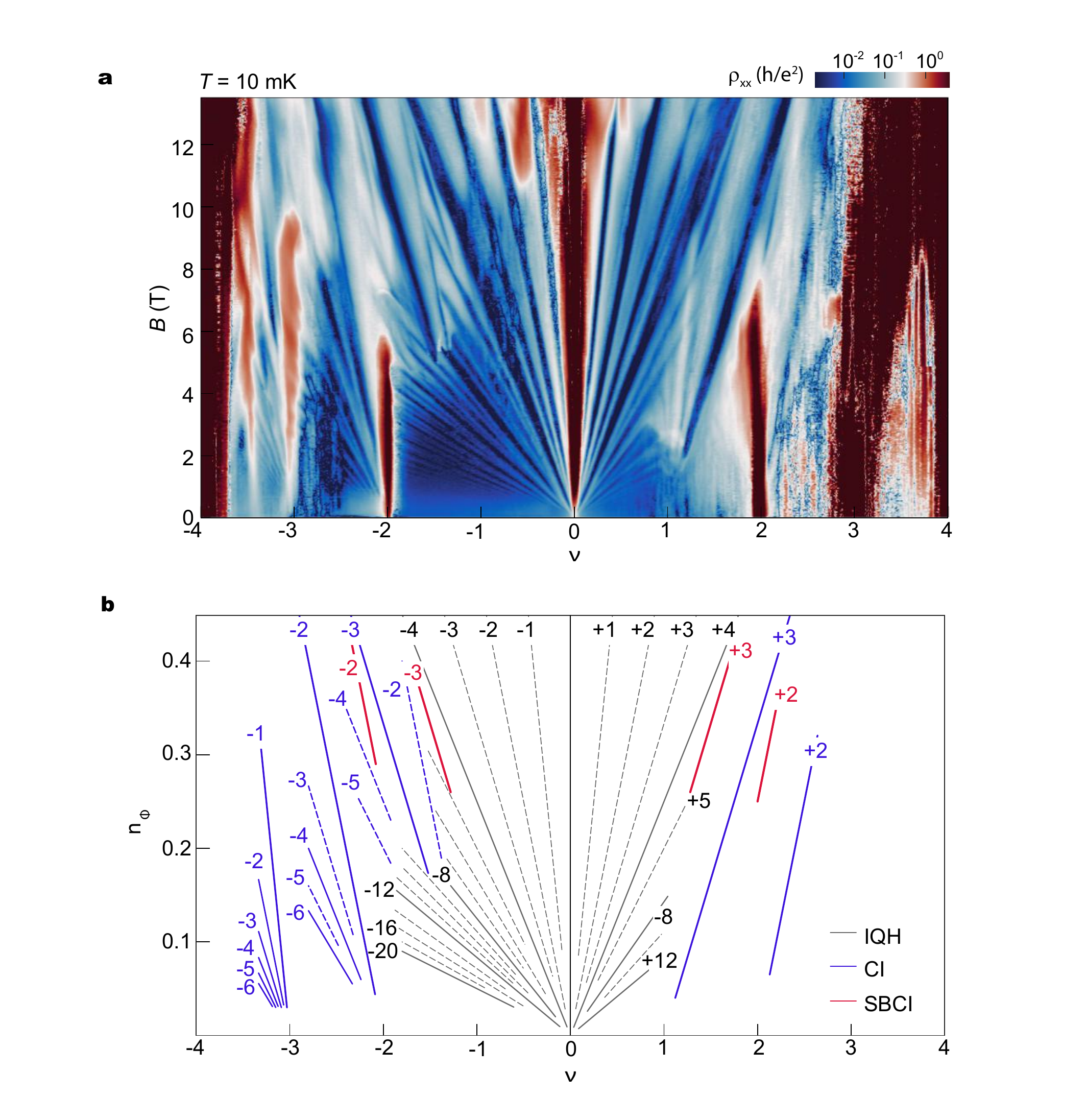}
\caption{\textbf{Landau fan diagram up to 13.5 T.}
 \textbf{a}, $\rho_\mathrm{xx}$ as a function of $\nu$ and $B$ at a nominal temperature of 10 mK.
 \textbf{b}, Schematic of observed Chern insulator structure, showing gapped states that follow linear trajectories parameterized by $\nu = tn_\mathrm{\Phi} + s$. $n_\Phi$ and $\nu$ are the magnetic flux quanta and number of electrons per moir\'e unit cell, respectively. 
 Three trajectory classes are distinguished by color: Integer quantum Hall (IQH) (gray; $s = 0, t \in \mathbb{Z})$, Chern insulators (CI) (blue; $s, t \in \mathbb{Z}, s \neq 0$) and symmetry broken Chern insulators (SBCI) (magenta; fractional $s$, $t \in \mathbb{Z}$). }
\label{fig:Full LL supple}
\end{figure*}

\begin{figure*}[ht!]
\includegraphics[width= 7.2in]{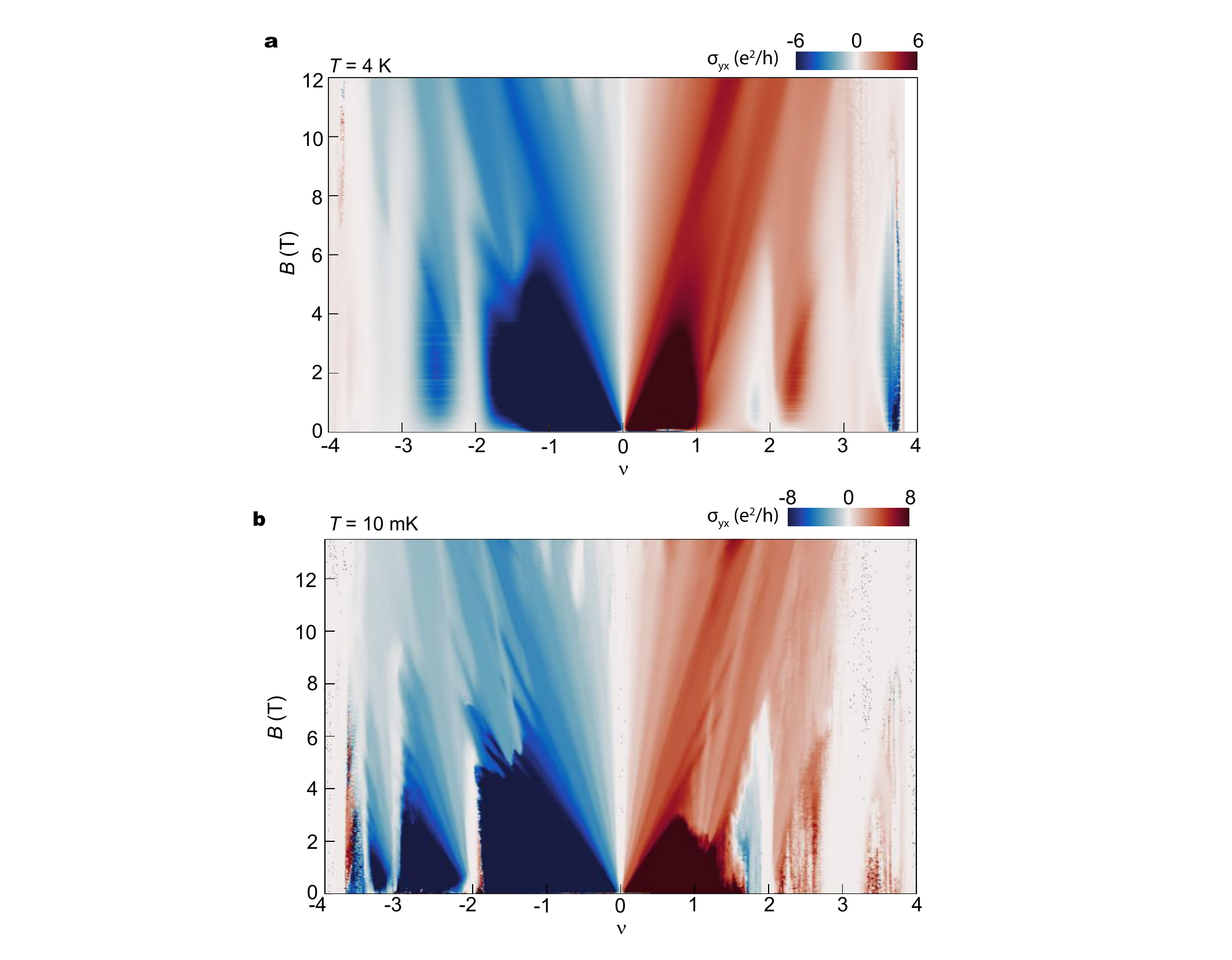}
\caption{\textbf{Landau fan diagram of Hall conductivity.}
\textbf{a, b}, $\sigma_\mathrm{yx}$ as a function of $\nu$ and $B$ at 4 K (\textbf{a}) and 10 mK (\textbf{b}).}
\label{fig:Hall_landaufan}
\end{figure*}

\begin{figure*}[ht!]
\includegraphics[width= 7.2in]{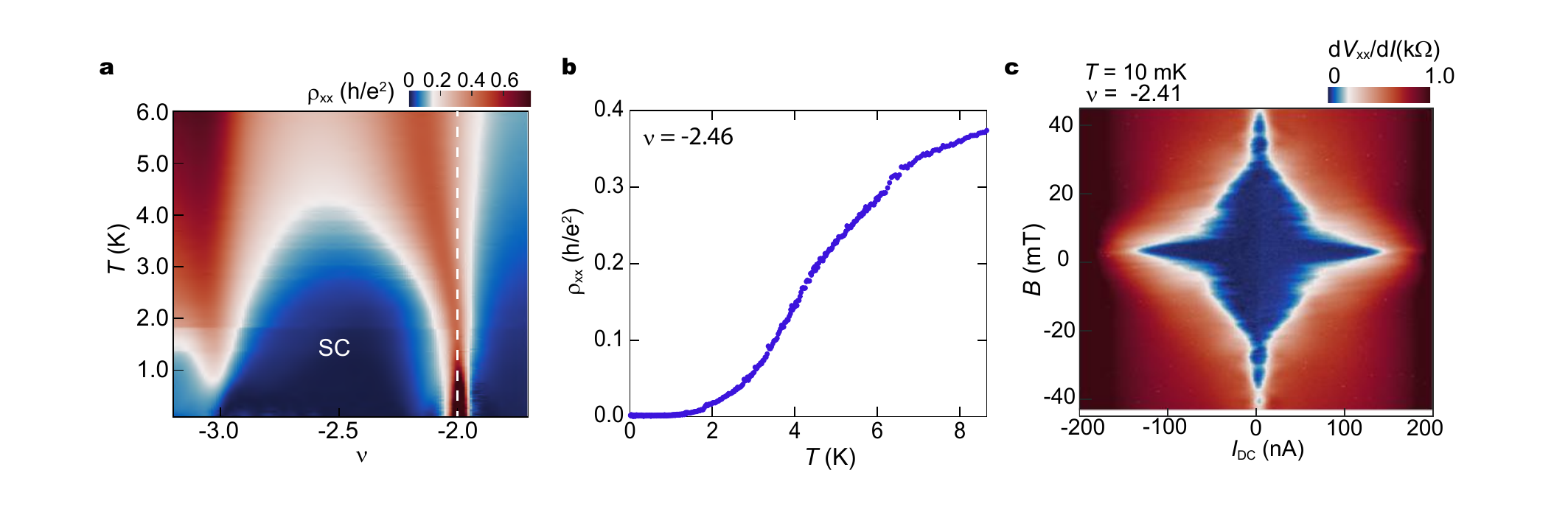}
\caption{\textbf{Superconductivity near $\nu = -2$.}
 \textbf{a}, $\rho_\mathrm{xx}$ as a function of $\nu$ and $T$ around $\nu = -2$.
 \textbf{b}, Line cut of $\rho_\mathrm{xx}$ as a function of $T$ at $\nu = -2.46$.
 \textbf{c}, $dV_\mathrm{xx}/dVI$ as a function of magnetic field at a nominal temperature of 10 mK and $\nu = -2$. Both current and magnetic fields suppress the low resistance state with Fraunhofer like oscillation.
 }
\label{fig:SC}
\end{figure*}

\begin{figure*}[ht!]
\includegraphics[width= 7.2in]{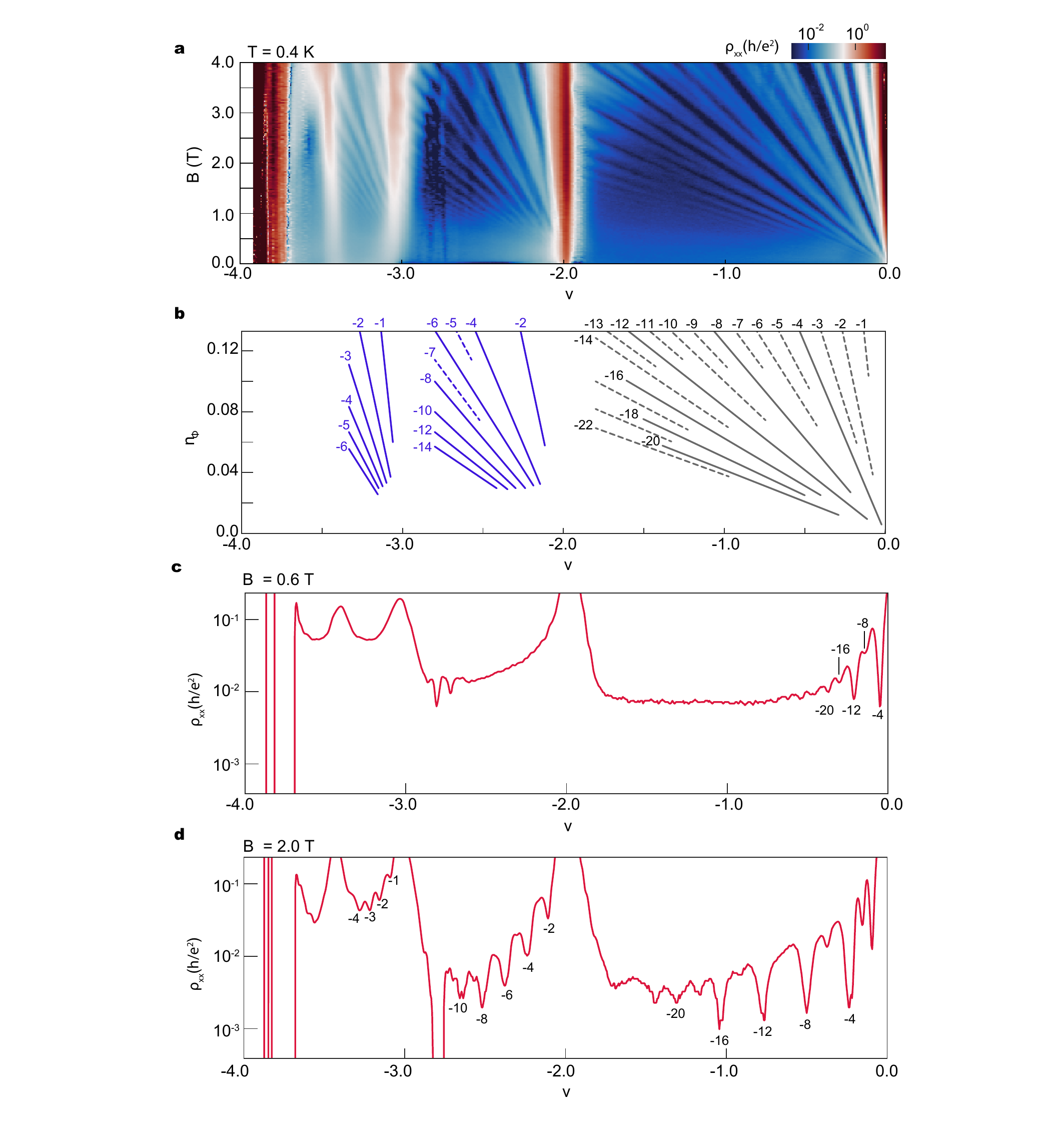}
\caption{\textbf{Landau fan diagram at low magnetic fields at 0.4 K.}
 \textbf{a}, $\rho_\mathrm{xx}$ as a function of $\nu$ and $B$.
 \textbf{b}, Schematic observed Chern insulator structure based on $\textbf{a}$.
 \textbf{c}, Line cuts of $\rho_\mathrm{xx}$ at 0.6 T in \textbf{a}.
 \textbf{d}, Line cuts of $\rho_\mathrm{xx}$ at 2.0 T in \textbf{a}.
 }
\label{fig:low field fan}
\end{figure*}

\begin{figure*}[ht!]
\includegraphics[width= 7.2in]{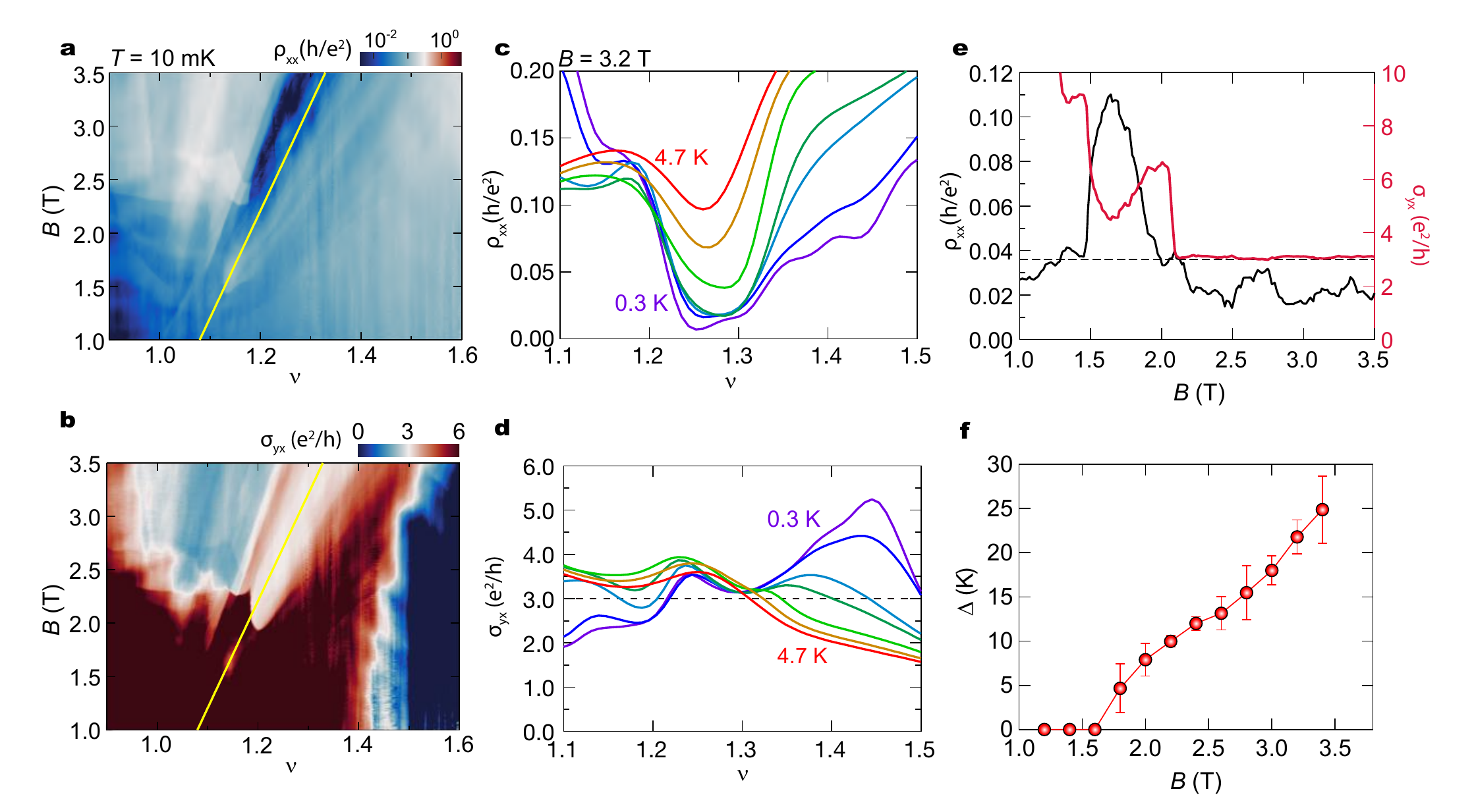}
 \caption{\textbf{$\bm{C = +3}$ Chern insulator from $\nu = +1$.}
\textbf{a, b}, $\rho_\mathrm{xx}$(\textbf{a}) and $\sigma_\mathrm{yx}$ (\textbf{b}) as a function of $\nu$ and $B$ in negative fillings at 10 mK.
 \textbf{c, d}, $\rho_\mathrm{xx}$(\textbf{c}) and $\sigma_\mathrm{yx}$ (\textbf{d}) as a function of $\nu$ in negative fillings between 0.3 and 4.7 K
 \textbf{e}, Line cuts of $\rho_\mathrm{xx}$(black) and $\sigma_\mathrm{yx}$ (red) as a function of $B$ along with yellow lines in \textbf{a} and \textbf{b}. Dashed line correspond to $3e^2/h$.
 \textbf{f}, The thermal activation gap of $C = +3$ Chern insulator as a function of $B$. The values are calculated from the fits to an Arrhenius law, $\rho_\mathrm{xx}\sim \exp(-\Delta /2T)$. Error bars in the gaps represent the uncertainty arising from determining the linear (thermally activated) regime for the fit.}
\label{fig:p_landaufan}
\end{figure*}

\begin{figure*}[ht!]
\includegraphics[width= 7.2in]{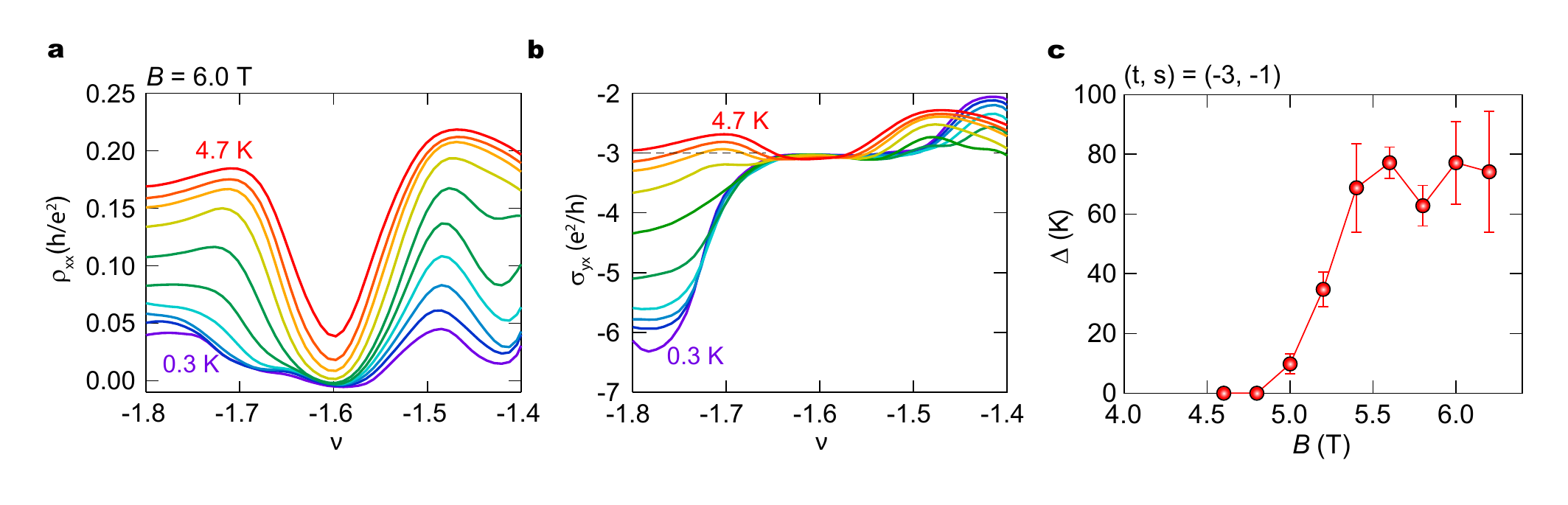}
 \caption{\textbf{Temperate dependence of a $\bm{C = -3}$ Chern insulator from $\nu = -1$.}
 \textbf{a, b}, $\rho_\mathrm{xx}$(\textbf{a}) and $\sigma_\mathrm{yx}$ (\textbf{b}) as a function of $\nu$ in negative fillings between 0.3 and 4.7 K.
\textbf{c}, The thermal activation gap of $C = -3$ Chern insulator as a function of $B$. The values are calculated from the fits to an Arrhenius law, $\rho_\mathrm{xx}\sim \exp(-\Delta /2T)$. Error bars in the gaps represent the uncertainty arising from determining the linear (thermally activated) regime for the fit.}
\label{fig:gap_temp}
\end{figure*}


\begin{figure*}[ht!]
\includegraphics[width= 7.2in]{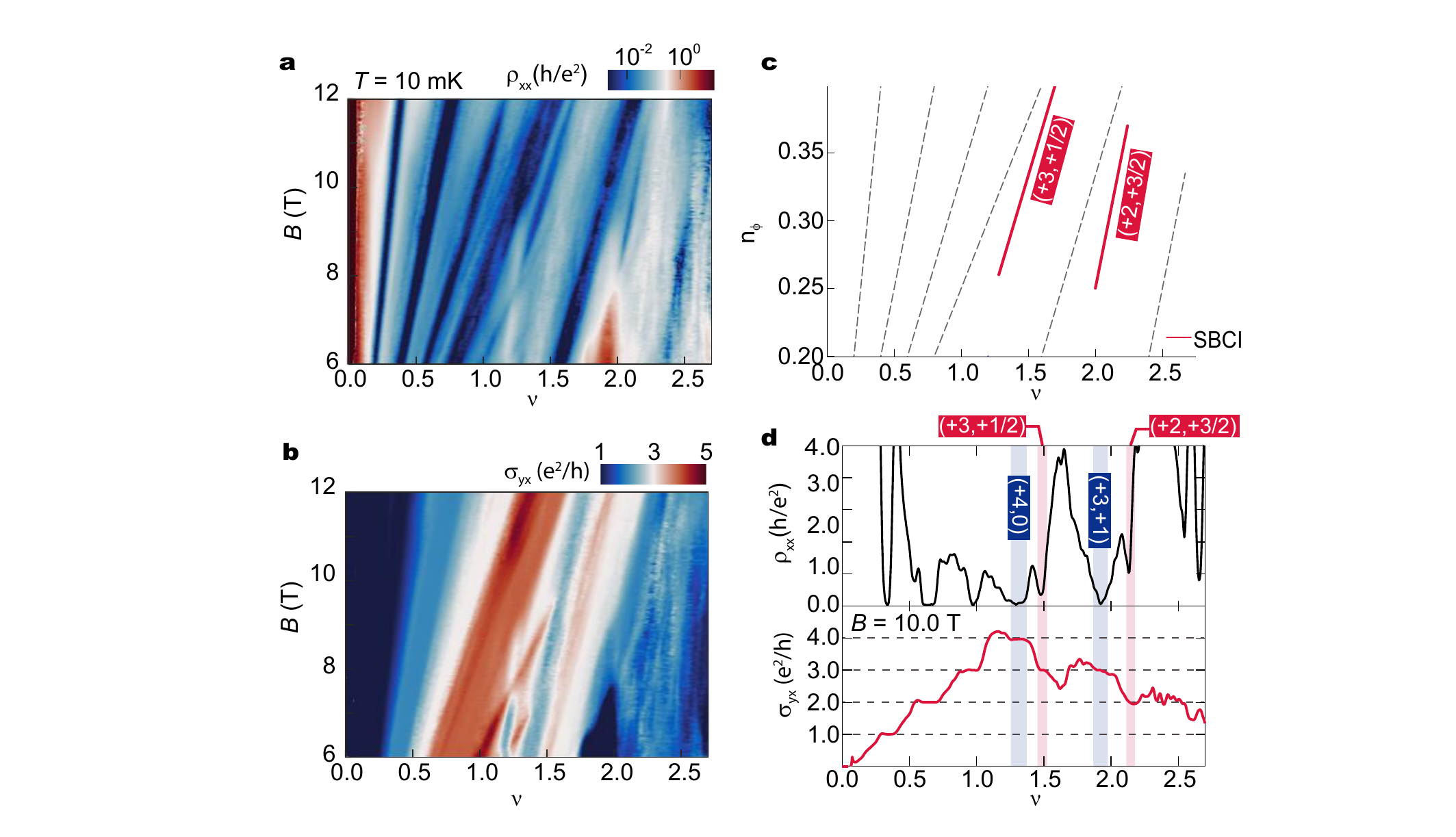}
 \caption{\textbf{Symmetry broken Chern insulators from $(t, s) = (+3, +1/2)$ and $(+2, +3/2)$.}
\textbf{a}, $\rho_\mathrm{xx}$ as a function of $\nu$ and $B$ in a range of 6-12 T.
\textbf{b}, $\sigma_\mathrm{yx}$ as a function of $\nu$ and $B$ in a range of 6-12 T.
 \textbf{c}, Schematic observed Chern insulators corresponding to the observations in \textbf{a} parameterized by $\nu = tn_\mathrm{\Phi} + s$. Red lines correspond to SBCI state with $(t, s) = (+3, +1/2)$ and $(+2,+3/2)$.
 \textbf{d}, Line cuts of $\rho_\mathrm{xx}$ and $\sigma_\mathrm{yx}$ around 10 T at positive fillings.}
\label{fig:sbci}
\end{figure*}

\begin{figure*}[ht!]
\includegraphics[width= 7.2in]{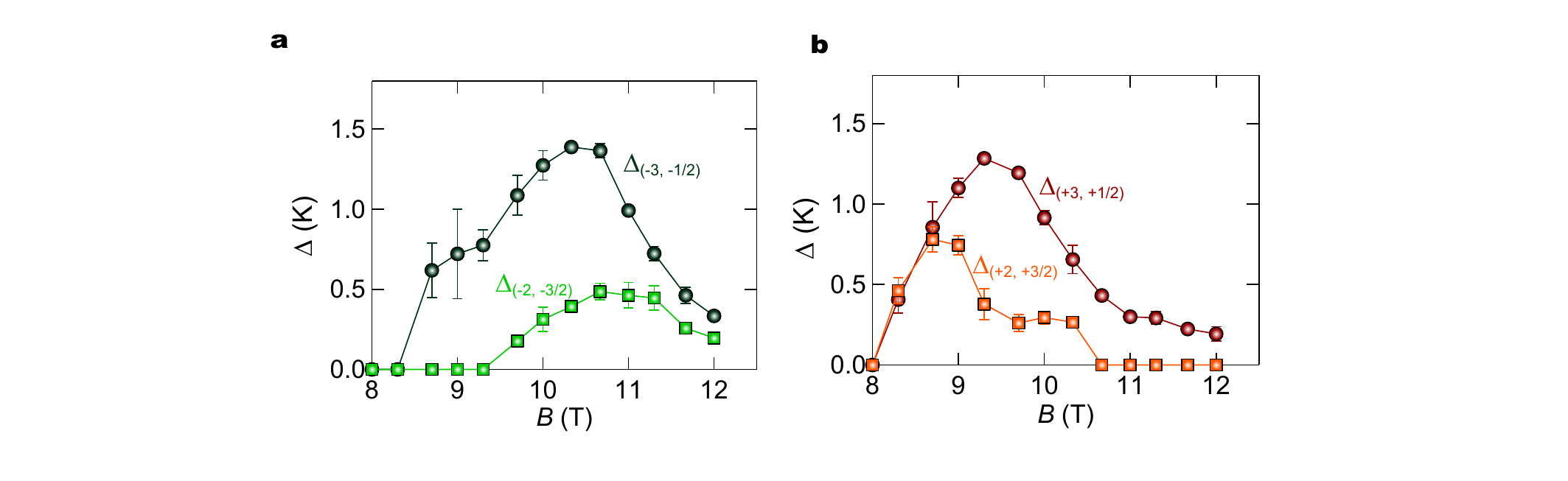}
 \caption{\textbf{Thermal activation gaps $\Delta$ of SBCIs.}
\textbf{a},  $\Delta$ of SBCI with $(t, s) = (-3, -1/2)$ and $(-2, -3/2)$ as a function of $B$.
\textbf{b},  $\Delta$ of SBCI with $(t, s) = (+3, +1/2)$ and $(+2, +3/2)$ as a function of $B$.
}
\label{fig:sbci_gap}
\end{figure*}

\clearpage

\subsection{Low field quantum oscillation analysis}
The low charge- and twist angle disorder of our device allows us to resolve quantum oscillations down to very low magnetic fields.  Figure \ref{fig:low field fan} shows quantum oscillation data for $-4<\nu<0$.  At a magnetic of 0.6 T, quantum oscillations are  not visible except originating from the charge neutrality point.  Notably, the strongest features are at Landau level filling $\nu_{LL}=\pm4, \pm12, \pm20$ (Fig. \ref{fig:low field fan}c), precisely as expected from the noninteracting band structure of twisted bilayer graphene\cite{bistritzer_moire_2011}. This sequence can be understood as twice that of monolayer graphene, and indeed arises from the two gapless Dirac cones of the constituent graphene monolayers.  

This is in marked contrast to published quantum oscillations in other twisted bilayers which uniformly show a sequence of $\pm4$, $\pm8$, $\pm12$, etc. However, we see the emergence of the more typical sequence at higher magnetic fields, for example in the data taken at $B = 2$ T shown in Fig. \ref{fig:low field fan}c. By this magnetic field, we also see quantum oscillations near $\nu=-2$ and $\nu=-3$, with sequences of $-2, -4, ,6,...$ and $-1,-2,-3,-4....$, respectively.  The transition in behavior may arise from a single particle effect such as strain\cite{zhang_landau_2019}, which may well be ubiquitous in graphene devices at some level.  However, conventional quantum Hall ferromagnetism may also be at play, as evidenced by the emergence of additional quantum oscillation minima at $\nu_\mathrm{LL}=1,2,3$ and $6$. 

The data can be elegantly explained under an ansatz consistent with that taken in Ref. \cite{zondiner_cascade_2020}, with the modification that no three-fold symmetry state exists.  In this picture, $-4<\nu<-3$ hosts a one component Fermi surface, $-3<\nu<-2$ has a two component Fermi surface, and $-2<\nu<0$ has a four component Fermi surface. A key conclusion from our data is that the failure to observe the expected sequences near $\nu=-2$ (where two copies of gapless Dirac, $\nu_\mathrm{LL}=-2,-6,-10$ would be expected) or $\nu=-3$ (where a single Dirac point is expected, $\nu_\mathrm{LL}=-1,3,-5$) are consistent with the breakdown of symmetry observed at neutrality at the magnetic fields where oscillations are observed in the other fans.  Cleaner samples, allowing measurements of the $\nu-2$ and $\nu=-3$ Landau fans, can be expected to similarly reveal the underlying Dirac physics.  

\subsection{Continuum model}

We calculated the zero-field dispersion and density of states using the continuum model of twisted bilayer graphene\cite{bistritzer_moire_2011-1,koshino_maximally_2018}. Given the large amount of literature written on this model, we give only a succinct summary here.

For a twist angle $\theta$ and the graphene lattice constant $a = 0.246$ nm, the distance between the ${\bf K}_\ell$ points of the two layers $\ell = 1,2$ in momentum space is $k_\theta = \frac{4 \pi}{3a} 2 \sin \theta/2$. The corresponding Moir\'{e} reciprocal lattice vectors are 
\begin{equation}
	{\bf G}^M_1 = k_\theta ( - \sqrt{3}, -3) /2, \; \; {\bf G}^M_2 = k_\theta (\sqrt{3},0). 
\end{equation}
Consequently, the Moir\'{e} lattice constant is $a_M = \frac{a}{2 \sin \theta/2}$. For the intralayer Hamiltonian in the ${\bf K}$-valley we take
\begin{equation}
	H_\ell ({\bf k}) = -v \left[ R(\pm \theta/2) ({\bf k} - {\bf K}_\ell) \right] \cdot {\bf \sigma}
	\label{Eq:IntraLayer}
\end{equation}
where $R(\pm \theta/2)$ is a rotation matrix by angle $\theta/2$ for layers $\ell =1,2$, respectively, and $\sigma=(\sigma_x,\sigma_y)$ are the Pauli matrices acting in the subspace of two graphene sublattices. We set the Dirac velocity to be $v = 2.51 a$ eV, where $a$ is the graphene lattice constant. We have taken $\hbar=1$. 

The interlayer Moir\'{e} coupling $H_M$ allows for scattering between neighboring Moir\'{e} Brillouin zones. It is common to write this Hamiltonian in real space. For the ${\bf K}$-valley, it is given by:
\begin{eqnarray}
	H_M ({\bf r}) &=& w \begin{pmatrix}
		\alpha & 1 \\ 1 & \alpha 
	\end{pmatrix} 
	+ w \begin{pmatrix}
		\alpha & e^{-i \phi} \\ e^{i \phi} & \alpha 
	\end{pmatrix}e^{i  {\bf G}^M_1 \cdot {\bf r}} 
	+ w \begin{pmatrix}
		\alpha &  e^{i \phi}\\ e^{-i \phi} & \alpha
	\end{pmatrix} e^{i  ({\bf G}^M_1 + {\bf G}^M_2) \cdot {\bf r}}
\end{eqnarray}
where $\phi = 2\pi/3$. Note that in momentum-space, a term of the form $e^{i  {\bf G}^M_1 \cdot {\bf r}} $ allows scattering from ${\bf k}$ to ${\bf k} + {\bf G}^M_1$. We set $w=110$ meV, and $\alpha = 0.8$ to model the lattice relaxation. 
We include Moir\'{e} mini-Brillouin zones around the ${\bf K}$-point of the monolayer graphene up to a cut-off distance of $\sim 5 |{\bf G}^M|$.

\subsection{Calculation of Hofstadter butterfly}

To calculate the effect of a finite magnetic field, we performed the Peierls substitution ${\bf k} \rightarrow {\bf k} - e {\bf A}$. The intralayer Hamiltonian Eq.~\eqref{Eq:IntraLayer} is diagonalized in a standard Landau level basis $| L \alpha n y \rangle$, where $L$ is the layer, $\alpha$ the sublattice, $n$ the Landau level index and $y$ the guiding center coordinate. We then write the interlayer coupling $H_M$ in this basis following Refs.~\cite{bistritzer_moire_2011,zhang_landau_2019,hejazi_landau_2019,lian_landau_2020}. The spectrum is calculated for flux per Moir\'{e} unit cell $\Phi = \Phi_0 \frac{p}{q}$ with $p,q$ coprime and $q\leq 30$. Note that the spectrum is the same for both spin and valley species, so we only compute it here for one valley, neglecting the Zeeman spin splitting.

\begin{figure}[ht!]
	\includegraphics[height=6cm]{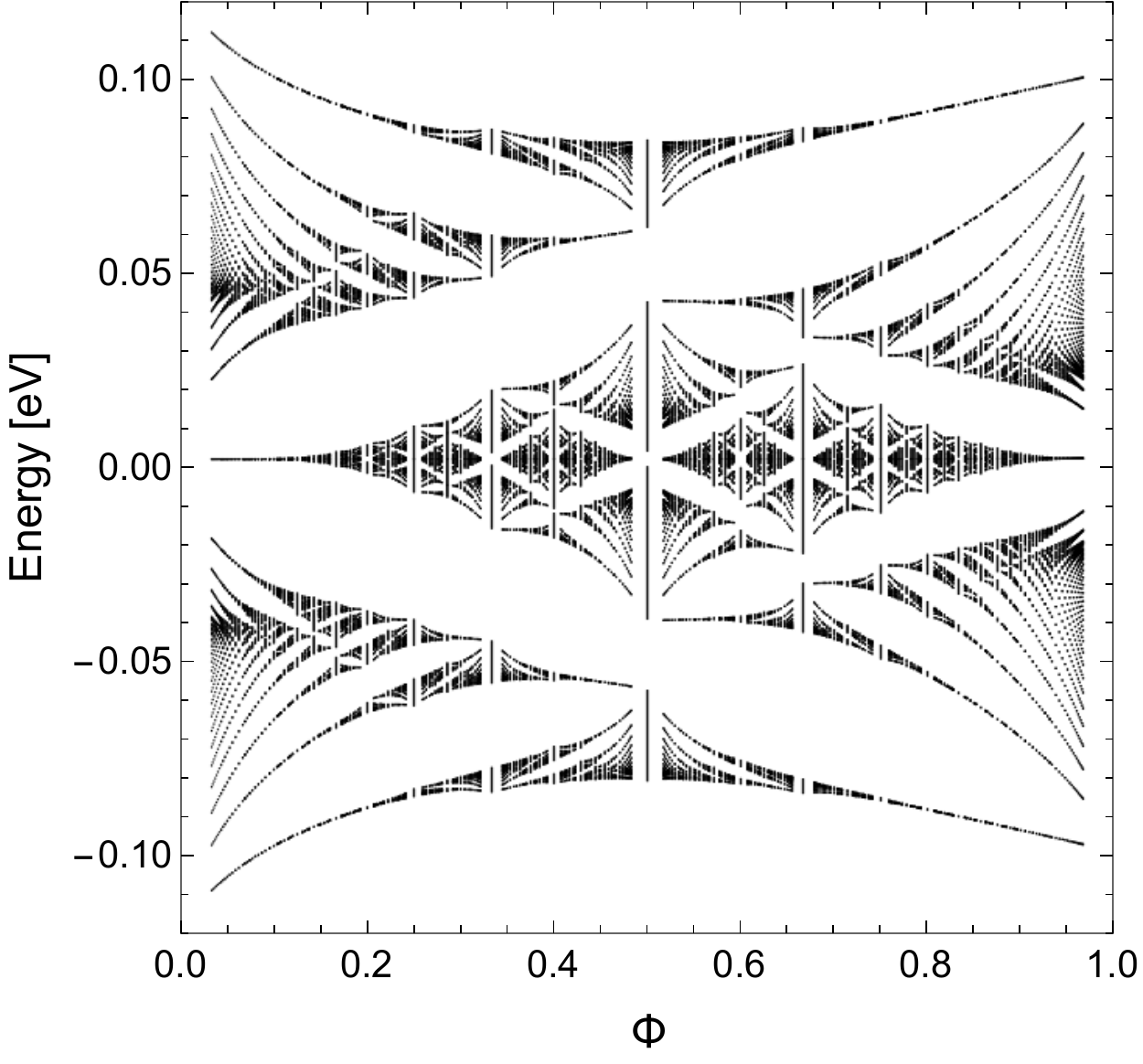}
	\caption{
	\textbf{The Hofstadter butterfly calculated for $\theta = 1.8^\circ$ and further parameters outlined in the text.} Each state pictured is fourfold (spin and valley) degenerate.}
	\label{Fig:SupplTheory}
\end{figure}

The resulting spectrum, known as the Hofstadter butterfly\cite{hofstadter_energy_1976}, turned out to match the experiments best for twist angle $\theta = 1.8^\circ$, see Fig.~\ref{Fig:SupplTheory}. In the literature (see Refs. \onlinecite{bistritzer_moire_2011,zhang_landau_2019,hejazi_landau_2019,lian_landau_2020}), there is a large variety of shapes of butterflies for the same twist angle. Further experimental study of the exact spectrum might elucidate which model parameters are best suited to describe tBLG. In order to fit the data, we have rescaled the energy in units of the bandwidth. The full Hofstadter butterfly is shown in Fig.~\ref{Fig:SupplTheory}.

\begin{figure}[ht!]
	\includegraphics[height=4cm]{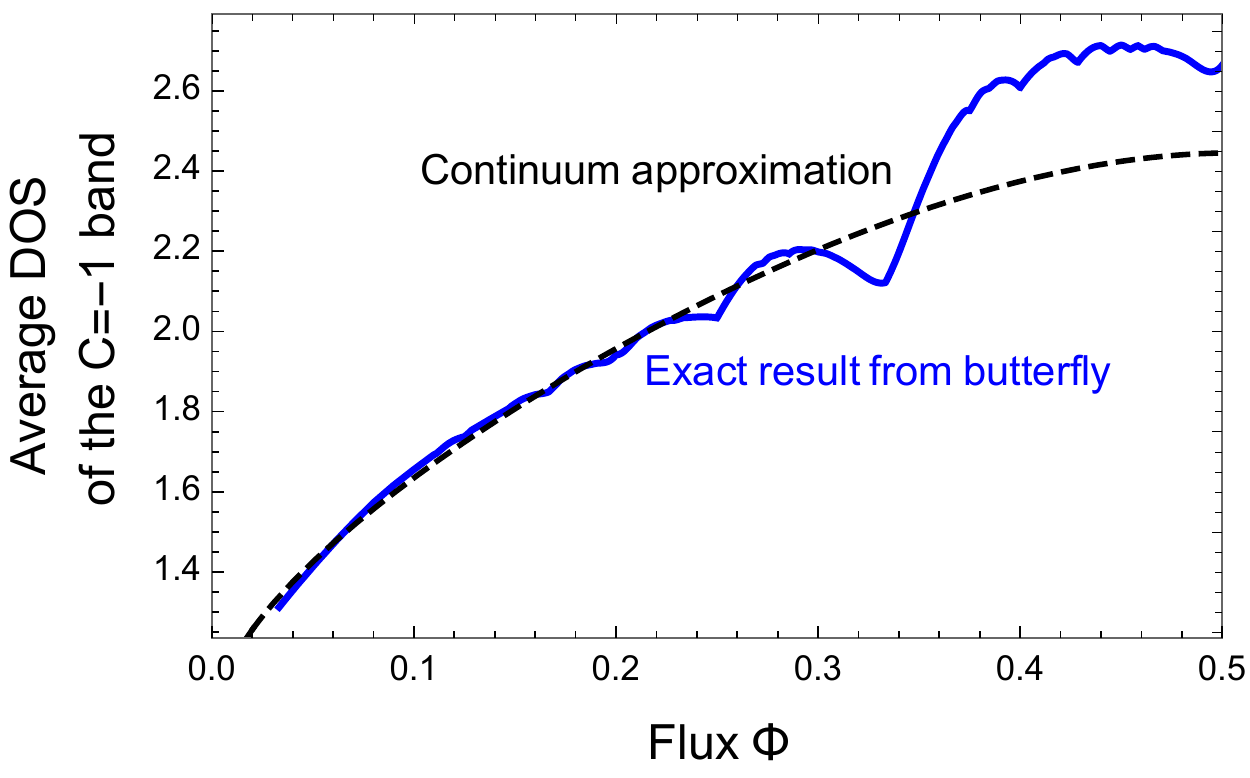}
	\caption{
	The average density of states of the $C=-1$ band is given by the number of states $n(\Phi) = 1 - \Phi$ divided by its bandwidth $W(\Phi)$. Since the bandwidth decreases faster than the number of states, the net density of states is an increasing function of flux. The continuum approximation here is given by taking the bandwidth $W_{-1}(\Phi) = W_0 - a v_F^* \sqrt{\Phi} - b \Phi / m^*$.
	\label{Fig:SupplTheoryDOS}}
\end{figure}

The dominant features of the Hofstadter butterfly are the band that evolves from the zeroth Landau level at small magnetic field, with Chern number +2 per spin/valley, flanked by two $C=-1$ bands to give a net zero Chern number for the fully filled band. The bandwidth of the $C=-1$ band at small fluxes can be approximated by the simple formula $W_{-1}(\Phi) = W_0 - a v_F^* \sqrt{\Phi} - b \Phi / m^*$, where $a,b$ are scaling constants, $v_F^*$ is the effective Dirac velocity at charge neutrality and $m^*$ is the effective mass at the flat band edges. Because of this scaling, the net density of states in the $C=-1$ band is increasing with flux, as can be seen in Fig.~\ref{Fig:SupplTheoryDOS}.  Consequently, for a fixed interaction strength one expects a first order Stoner transition into a spin/valley polarized state as a function of $B$, if one does not already exist at $B=0$.

\subsubsection{Disorder}

In the absence of disorder, Landau levels are perfectly degenerate and are therefore always susceptible to quantum Hall ferromagnetism. Disorder, however, will broaden the Landau levels. Any realistic assessment of possible symmetry breaking in Landau levels requires a study of the role of disorder.

In tBLG, the dominant form of disorder is twist angle inhomogeneity~\cite{wilson_disorder_2020}. This mostly affects the band edges, as the bandwidth of the flat bands is highly sensitive to twist angle. Here, we model the disorder phenomenologically by add a self-energy that is increasing with energy away from charge neutrality, $\mathrm{Im} \Sigma (\epsilon) = - \gamma_B | \epsilon|$, where $\gamma_B$ is a dimensionless parameter proportional to the strength of disorder. In the final calculations we set $\gamma = 0.001$. Note that this shape of disorder self-energy also applies to single-layer graphene in the presence of short-range potential impurity scattering.

Using this disorder potential, we can convert the Hofstadter spectrum into a density of states $\rho (\Phi, \epsilon)$, which is used in Fig. 1 of the main text.

\subsubsection{Hartree-Fock description of interaction effects}
\label{Sec:HF}
To account for the state with spontaneous polarization, we performed a simplified Hartree-Fock calculation for each flux.As free parameters we introduce separate chemical potentials $\mu_{\sigma \xi}$ for each spin $\sigma = \uparrow, \downarrow$ and valley $\xi = \pm1$. The filling per spin/valley is
\begin{equation}
	\nu_{\sigma \xi} = \int_0^{\mu_{\sigma \xi}} d\epsilon \; \rho(\epsilon),
\end{equation}
where $\rho(\epsilon)$ is the density of states obtained from the Hofstadter butterfly described above.
The total kinetic energy is given by
\begin{equation}
	K = \sum_{\sigma \xi} \int_0^{\mu_{\sigma \xi}} d\epsilon \; \epsilon \; \rho(\epsilon).
	\label{HFKinetic}
\end{equation}
The total interaction energy is a Hartree-Fock approximation of a generic repulsion between the spin and valleys\cite{zondiner_cascade_2020},
\begin{equation}
	V = U \sum_{(\sigma \xi) \neq (\sigma' \xi') } \nu_{\sigma \xi} \nu _{\sigma' \xi'}
	\label{HFInteraction}
\end{equation}
where $U$ denotes the strength of the interactions. We minimized the total energy $K+V$ by varying the chemical potentials $\mu_{\sigma \xi}$, with the constraint that the total filling $\sum_{\sigma \xi} \nu_{\sigma \xi} = \nu$ is constant. We repeated this calculation for a range of densities $\nu \in [-4,4]$ and fluxes $\Phi \in [0,1/2]$. 

\begin{figure}
	\includegraphics[width=0.3\textwidth]{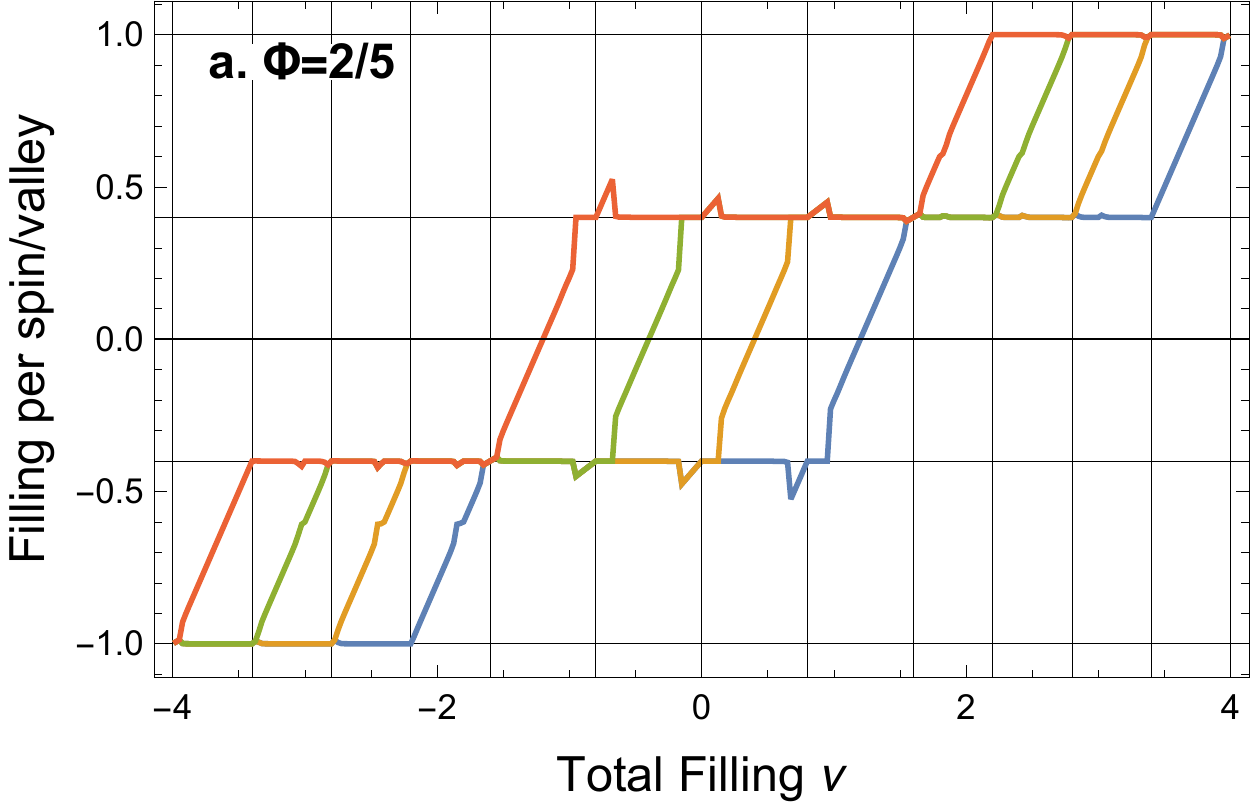}
	\includegraphics[width=0.3\textwidth]{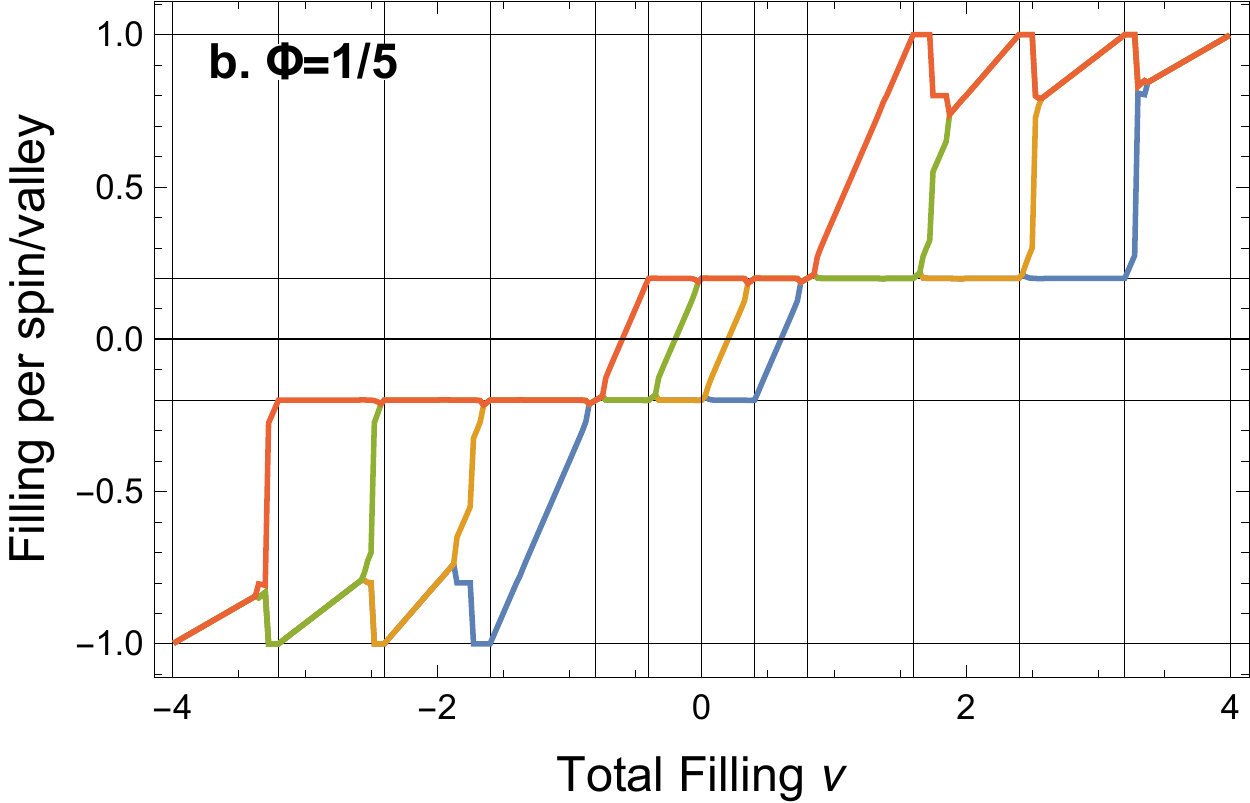}
	\includegraphics[width=0.3\textwidth]{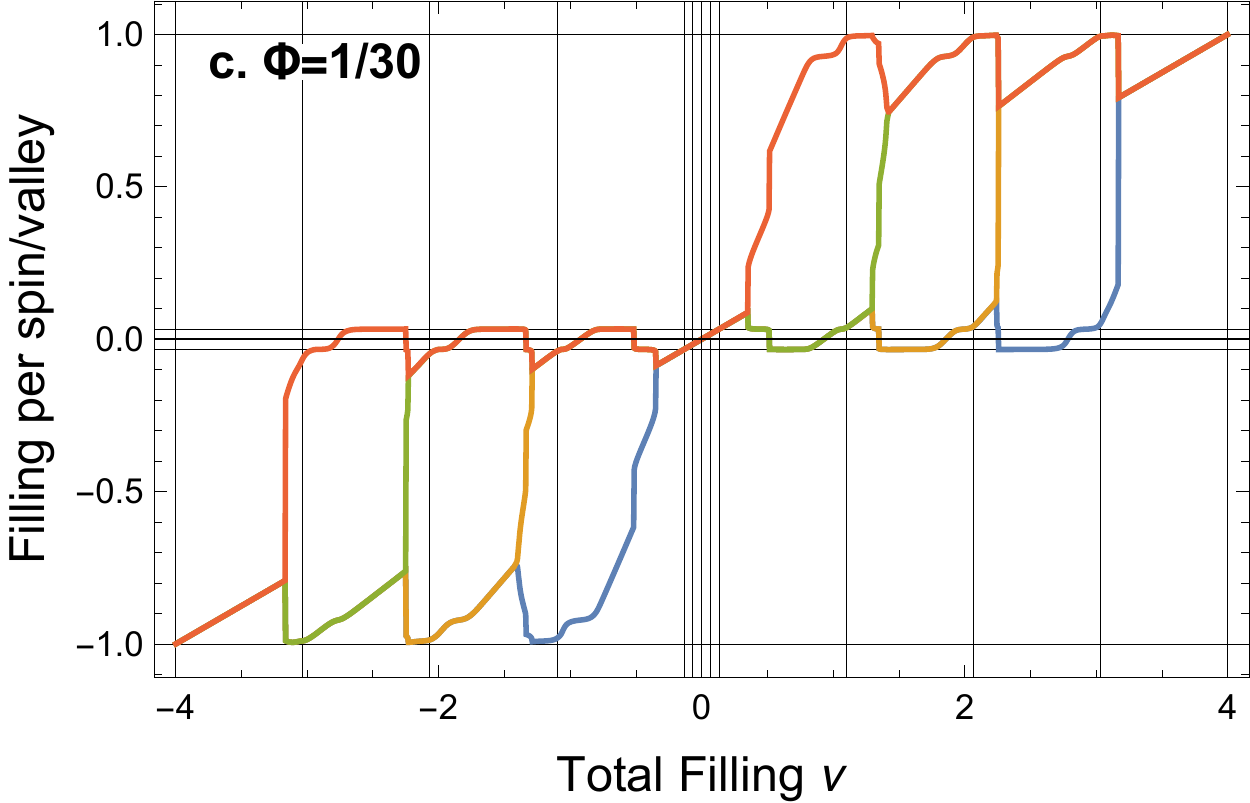}
	\caption{
	\label{Fig:SupplTheory2}
    Filling per spin/valley as a function of total filling, obtained using Hartree-Fock calculations with $U=0.41W$ for three typical values of the magnetic flux $\Phi = 2/5$, $1/5$ and $1/30$. At low fluxes, the sequence of fillings is similar to the zero-field result measured in Ref.~\cite{zondiner_cascade_2020}, in which the filling per spin/valley is reset to charge neutrality in a cascade of transitions. At a higher magnetic field a similar cascade occurs, but now the filling per spin/valley is reset to the $|C|=1$ gap. This leads to the sequence of Chern insulator states with $|C| = 1,2,3,4$ observed in the experiments.}
\end{figure}

In Fig.~\ref{Fig:SupplTheory2} we show the resulting filling per spin/valley as a function of total filling, for three values of the flux: $\Phi = 2/5$, $\Phi = 1/5$ and $\Phi = 1/30$, and $U=0.41W$. In the absence of spontaneous polarization, each spin/valley would be equally occupied. This always occurs when the total filling is $\nu = \pm 4 \Phi$, when all spin/valley degrees of freedom have fully occupied or unoccupied the $C=-1$ Hofstadter miniband.

For small fluxes (Fig.~\ref{Fig:SupplTheory2}b and c), the spin/valley polarization follows the same pattern as proposed by Ref.~\cite{zondiner_cascade_2020}. Starting at $\nu=-4$, all four spin/valley flavors increase their filling in parallel, when a first order transition occurs that fixes one of the spin/valley flavors to a filling $\nu = -\Phi$, and at the same time 'resets' the other three fillings to $\nu = -1$. This process of parallel filling followed by a reset continues until the $C=-1$ band is completely filled.

In contrast, at the higher flux $\Phi=2/5$ the filling of the $C=-1$ band occurs sequentially: first one spin/valley flavor completely fills the $C=-1$ band, followed by the second flavor, and so forth. This is a familiar pattern of ferromagnetic states, which occurs, e.g. in the interaction-split zeroth Landau level in graphene.

Note that this sequential filling at higher fluxes is consistent with the experimental results. For example, in between the (-3,-1) and (-4,0) state, one finds a (-2,-1) state. This state can be understood by having three spin/valley degrees of freedom completely filling the $C=-1$ band, whereas one spin-valley flavour partially fills the $C=-1$ band up to the $C=+1$ gap (see Fig.~\ref{fig:3})

\end{document}